\documentclass[aps,prd,twocolumn,showpacs,superscriptaddress,groupedaddress]{revtex4}

\usepackage{graphicx}

\usepackage{dcolumn}
\usepackage{bm}

\usepackage{textcomp} 
\usepackage{amstext} 

\usepackage{float}
\usepackage[T1]{fontenc}
\usepackage[utf8]{inputenc}

\usepackage[utf8]{inputenc}
\usepackage{graphicx}
\usepackage{epsfig}
\usepackage{mathtools, nccmath}
\usepackage{environ}
\usepackage{tabularx}
\usepackage[usenames]{color}

\NewEnviron{myequation}{%
\begin{equation}
\scalebox{0.85}{$\BODY$}
\end{equation}}

\usepackage[colorlinks,filecolor=blue,citecolor=blue,urlcolor=blue]{hyperref}
\usepackage[dvipsnames]{xcolor}
 
\hypersetup{pdfstartview=FitH, linkcolor=Blue,urlcolor=Blue, colorlinks=true}

\RequirePackage{lineno}

\begin{document}

\title{A possible search for Majorana neutrinos at future lepton colliders}


\author{E.~Antonov}
\author{A.~Drutskoy}
\affiliation{P.N. Lebedev Physical Institute of the Russian Academy of Sciences, Moscow 119991, Russia}

\author{M.~Dubinin}
\affiliation{Skobeltsyn Institute of Nuclear Physics (SINP MSU), M.V. Lomonosov Moscow State University, Leninskie gory,
GSP-1, 119991 Moscow, Russia}

\date{\today}

\begin{abstract}
\vspace*{0.01cm}
We discuss the process $\ell^+\ell^- \to N W^{\pm} \ell^{\mp}$,
where $N$ is a heavy Majorana neutrino and $\ell = e, \mu$.
Large cross sections are expected for these processes at high center-of-mass energies,
which can be reached at future lepton-lepton colliders.
The Monte Carlo simulation of the studied processes is produced within the framework of the seesaw type-I model,
where the Majorana neutrinos (or heavy neutral leptons, HNL), are introduced in the standard leptonic sector.
Recently the possibility to search for the direct HNL production was studied in the $\ell^+\ell^- \to N \nu_{\ell}$ process
with the subsequent decay $N \to W^{\pm} \ell^{\mp}$.
In this paper we investigate 
an alternative process $\ell^+\ell^- \to N W^{\pm} \ell^{\mp} \to W^{\pm}W^{\pm}\,\ell^{\mp}\,\ell^{\mp}$
with the lepton number violation by two units.
The similar processes appear in collisions with the same-sign beams,
$e^-e^- \to N W^-\,e^- \to W^-\,W^-\,e^+\,e^-$ or $\mu^+\mu^+ \to N W^+\,\mu^+ \to W^+\,W^+\,\mu^+\,\mu^-$.
The cross sections of the processes under consideration are enhanced by the soft photon exchange in the $t$-channel. We calculate the cross sections
for the signals and potential Standard Model backgrounds for
the $e^+e^-$ beam collisions at the 1 TeV center-of-mass energy and the $\mu^+\mu^-$ collisions at 3 TeV and 10 TeV.
Due to the diagrams with soft $t$-channel photons and respective interference the promptly emitted leptons are produced in the direction close to the corresponding beam.
These leptons will be lost in the beam pipe or badly measured by forward detectors.
However, the signal events can be well separated from backgrounds using
the rest of the event containing the $WW\ell$ particles.
Finally, the expected upper limits on the mixing parameters $|V_{\ell N}|^2$ as a function of M($N$) are calculated.
\end{abstract}

\smallskip





\maketitle

\section{\label{sec:intro}Introduction}

Neutrinos of the Standard Model (SM) can be Majorana fermions, origin of their masses is associated with a violation of $U(1)_{B-L}$ global invariance of the SM Lagrangian. As a result, very rare interaction processes with a violation of the lepton number (LNV processes) may be possible. 
Very small masses of the standard (or active) neutrinos are explained within the framework of extensions of the SM leptonic sector by introducing of
Majorana neutrinos (or heavy neutral leptons, HNL). In the processes involving HNL the violation of lepton number can be experimentally observed. 

The current experimental upper limits on the HNL mixing parameters $|V_{lN}|, \, l=e,\mu,\tau$ as a function of the HNL mass
are summarized in \cite{hnl_status}.
The mass interval from 0.1 GeV to a few GeV is covered by the extracted beam experiments with missing energy
reconstruction and the experiments with displaced vertices. The mass interval up to TeV scale is a range
for collider experiments. Rather strict upper limits
have been obtained at LEP for the HNL masses up to the $Z$ bozon mass \cite{reviewa}, however,
the limits on the HNL masses higher than 100~GeV were rather weak. A strong limit for the squared
sum $|\sum_i {V_{e N_i}^2 / M(N_i)}| \leq 1.8\times 10^{-8}$~GeV$^{-1}$ has been
obtained within the seesaw type-I scenario from the neutrinoless double beta ($0\nu\beta\beta$) decays~\cite{betabeta}.
However this limit can be circumvented in case of the three generations of the heavy HNL and 
a specific combination of $CP$ phases in the PMNS matrix~\cite{wwa}. Updated experimental upper limits on the mixing parameters for the HNL 
masses larger than 100~GeV have been provided by the LHC run II data~\cite{atlas,cmsa,cmsb}, better limits
are expected with the future LHC luminosity upgrade ~\cite{reviewa,ppa,ppb,ppc}.
A significant improvement of these upper limits can be reached at future high energy lepton
colliders~\cite{nne}.

Simulation of the HNL production in collider experiments demonstrated realistic possibilities of the signal searches
in the $e^+ e^-$, $e^- e^-$, $\mu^+ \mu^-$ and $\mu^+ \mu^+$ beam colliders at TeV energies.
After the partonic level calculation of the process $e^+ e^- \to q \bar q e \nu $ at LEP2 \cite{almeida} performed for complete set of $2\to 4$ diagrams by means of \mbox{CompHEP~\cite{comphep, comphepb}}, a number of detailed studies for the signal separation from the SM backgrounds including detector simulation has been performed. In \cite{nne} a study
for the channel $e^+ e^- \to N  \nu \to q \bar q l \nu $ at ILC and CLIC energies by the chain of the
FeynRules \cite{feynrules}, Whizard \cite{whizard} and Delphes \cite{delphes} packages is presented.
A field-theoretic model is defined in the FeynRules package,
event generation is done by Whizard. Detector simulation and event reconstruction is performed by Delphes,
thereby providing a realistic simulation of all final state objects. 
Potential to search for HNL in the $\mu^+ \mu^- \to N  \nu \to q \bar q l \nu $ process at TeV energies
has been studied in \cite{Felix} and \cite{kwok} using the FeynRules, Whizard, and Delphes chain and in \cite{li}
using the FeynRules, Madgraph5 \cite{madgraph}, and Delphes chain.
In these studies of the $\ell^+ \ell^- \to N  \nu \to q \bar q \ell \nu$ processes
the upper limits for the mixing parameter are obtained to be $|V_{\ell N}|^2 \sim 10^{-5} - 10^{-7}$
at integrated luminosities of about a few ab$^{-1}$ 
and the collision CM energy in the \mbox{(1--10)~TeV} range. 
Although background contributions under the signals are large, a good signal separation was
achieved by using a set of discriminating variables calculated in the final state.
The HNL production via the vector boson scattering process at high energy muon colliders has been studied in~\cite{tong}.
An opportunity to search for HNL at $e^- p$ collisions with TeV protons
and 60 GeV electrons has been discussed at \cite{gu}.

The production of the heavy Majorana neutrinos in the $e^+e^-\,\to\,N W^+e^-$ and $\mu^+\mu^-\,\to\,N W^+\mu^-$
processes is studied in this paper using the chain of the \mbox{LanHEP \cite{lanhep}}, CompHEP, Pythia6~\cite{pythia}, and Delphes packages.
A field-theoretic model is defined in LanHEP package, event generation is done by CompHEP,
the decays of HNL and \mbox{$W$ bosons} and hadronization are executed by Pythia, and detector simulation is performed by Delphes.
The produced Majorana neutrinos decay in the $N \to W^+ e^-$ and $N \to W^+ \mu^-$ modes
resulting in the lepton number violation by two units.
These processes are permitted only in case of the Majorana neutrinos and
have a clear experimental signature.
The charge-conjugated processes are assumed by default everywhere in the paper.
Similar processes $e^-e^- \to N W^- e^-$ and $\mu^+\mu^+ \to N W^+ \mu^+$ are
possible in the same-sign beam collisions. 

The $e^+e^-\,\to\,N W^+e^-$ process was studied
in~\cite{bane} at the CM energies 350 GeV and 500 GeV using the chain of the FeynRules
and Madgraph5 packages. However all final state particles were
required to be reconstructed in this study, that resulted in a very small number of the expected signal events.

It has to be noted, that the direct production of HNL at lepton-lepton collisions provides
a high sensitivity to the mixing parameter $|V_{\ell N}|^2$, because the process contains only one $NW\ell$ vertex.
The process $\mu^+\mu^+ \to W^+W^+$ with two such vertices gives worser upper limits, however 
has a sensitivity to the HNL masses higher than the CM energy~\cite{yang,wwa,wwb}.



\vspace{-0.3cm}
\section{\label{sec:exp}Seesaw type-I model}
\vspace{-0.1cm}

Neutrino mass generation using a seesaw type-I mechanism is performed by means of a Lagrangian~\cite{lagr}
\vspace{-0.2cm}
\begin{equation}
        \mathcal{L} = \mathcal{L}_{SM} + i \overline{\nu}_R \gamma^\mu \partial_\mu  \nu_R- \left( F \; \overline{L}_l  \tilde{H}\nu_R + \frac{1}{2} M_M \overline{\nu^c}_R \nu_R + h.c \right),
        \label{lagr}
\end{equation}
where
$L_l=(\nu_l, \, l)^T_L$ is the left lepton doublet, 
$\nu_R$ are HNL flavor states, $(\nu_R)^c \equiv C \overline{\nu}_R^T$ ($C=i \gamma_2 \gamma_0$),  
$H$ is the Higgs doublet
($\tilde{H}=i\tau_2 H^{*}$, $\tau_2$ is Pauli matrix),
$F$ is the Yukawa matrix 3$\times$3 and $M_{M}$ is a Majorana mass matrix, $M^T_M=M_M$. After spontaneous symmetry breaking the Yukawa matrix $M_{D}= F v /\sqrt{2}$ ($v=246$ GeV) and the Majorana mass matrix $M_M$ form a complete 6$\times$6 mass matrix in the extended lepton sector
\vspace{-0.1cm}
\begin{eqnarray} \label{M66}
    \frac{1}{2}(\overline{\nu}_L \vspace{2mm} \overline{\nu^c}_R)
        \mathcal{M}
    \left(
        \begin{array}{c}
        \nu_L^c \\
        \nu_R
        \end{array}
        \right)    +h.c. \, = \,
         && \\ \nonumber
         \frac{1}{2} (\overline{\nu}_L \vspace{2mm} \overline{\nu^c}_R) \left(
        \begin{array}{cc}
         0 & M_D\\
         M_D^T & M_M  
        \end{array}
        \right)
        \left(
        \begin{array}{c}
        \nu_L^c \\
        \nu_R
        \end{array}
        \right) +h.c.,
\end{eqnarray}
where the flavor states $(\nu_{L})_\alpha, (\nu_{R})_I$ and the mass states $\nu_k, N_I$ ($\alpha = e,\mu,\tau$, $k,I=1,2,3$) are connected by the transformation
\begin{eqnarray}
    \left(
    \begin{array}{c}
    \nu_L \\
    \nu_R^c 
    \end{array} \right)
    =\mathcal{U} P_L \left(
    \begin{array}{c} 
    \nu \\
    N 
    \end{array}
    \right),
    \hskip 5mm \mathcal{U}= \mathcal{W} \cdot {\rm diag}(U_\nu, U_N^*) 
\end{eqnarray}
$P_L=(1-\gamma_5)/2$, and $U_\nu, U_N$ are unitary $3 \times 3$ matrices. Block-diagonal form of the mass matrix \eqref{M66} looks as
\begin{eqnarray*}
    \mathcal{U}^\dagger \mathcal{M} \mathcal{U}^*
    =
    \left(
    \begin{array}{cc}
     U_\nu^\dagger & 0 \\
     0 & U_N^T  
    \end{array}
    \right) 
    \mathcal{W}^\dagger \mathcal{M} \mathcal{W}^*
    \left(
    \begin{array}{cc}
     U_\nu^* & 0 \\
     0 & U_N  
    \end{array}
    \right)
    \\
    =\left(
    \begin{array}{cc}
     U_\nu^\dagger m_\nu U_\nu^* & 0 \\
     0 & U_N^T M_N U_N  
    \end{array}
    \right)
    =
    \left(
    \begin{array}{cc}
     \hat{m} & 0 \\
     0 & \hat{M}  
    \end{array}
    \right),&&
\end{eqnarray*}
where
$\hat{m}$ and $\hat{M}$ are diagonal mass matrices for active neutrinos and HNL, $\mathcal{W}^\dagger \mathcal{M} \mathcal{W} = diag(m_\nu, \; M_N)$, $\mathcal{M}$ is defined by Eq.\eqref{M66}.
In the following diagonalization procedure \cite{ibarra1,ibarra2} the unitary $\mathcal{W}$-matrix is represented as an
exponent of an antihermitian matrix
\vspace{-0.1cm}
\begin{equation} 
\label{ow}
    \mathcal{W} = \exp \left(
    \begin{array}{cc}
        0 & \theta \\
        -\theta^\dagger & 0
    \end{array}\right)    
\end{equation}
and decomposed to second-order terms by $\theta$. As a result, we obtain a connection of mass states and
flavor states in the following form
\vspace{-0.1cm}
\begin{eqnarray*}
    \nu_L & \simeq & \left( 1- \frac{1}{2} \theta \theta^\dagger \right) U_\nu P_L \nu+ \theta U_N^* P_L N, 
    \label{nuL-3} 
    \\
    \nu^c_{R} & \simeq & -\theta^\dagger U_\nu P_L \nu + \left(1-\frac{1}{2} \theta^\dagger \theta \right) U_N^* P_L N.
\end{eqnarray*}
The first term in \eqref{nuL-3} corresponds to the the well-known phenomenological relation which defines the PMNS mixing matrix $\nu_{L \alpha} = \sum_{\alpha} (U_{\rm PMNS})_{\alpha j} P_L \nu_{j}$ \cite{pmns,pmnsb}. Deviation from unitarity for the PMNS matrix is
defined by the term $-\frac{1}{2}\theta \theta^\dagger$.  The Lagrangian terms for neutrino mass states and HNL mass states interaction with $W^\pm, Z$ bosons have the form
\begin{eqnarray} \label{ncurr}
    \mathcal{L}_{NC}^\nu &=& - \frac{g}{2 c_W} \gamma^\mu \overline{\nu}_L U_{\rm PMNS}^\dagger U_{\rm PMNS} \nu_L Z_\mu,\\  \nonumber
    \mathcal{L}_{CC}^\nu &=& -\frac{g}{\sqrt{2}}\overline{l}_L \gamma^\mu  U_{\rm PMNS} \nu_L W_\mu^- + h.c.,\\ \nonumber
    \mathcal{L}_{NC}^N &=& -\frac{g}{2 c_W} \overline{N}_L \gamma^\mu U_N^T \theta^\dagger \theta U_N^* N_L Z_\mu\\ \nonumber
    &-& \left[ \frac{g}{2 c_W} \overline{\nu}_L \gamma^\mu U_\nu^\dagger \theta U_N^* N_L Z_\mu +h.c. \right],\\ \nonumber 
    \mathcal{L}_{CC}^N &=& -\frac{g}{\sqrt{2}} \overline{l}_L \gamma^\mu \theta U_N^* N_L W_\mu^-+ h.c.
\end{eqnarray}
HNL mixing is defined in the approximation $\mathcal{W} \sim \mathcal{O}(\theta^2)$ as  
$\Theta  \equiv  \theta U_N^*$. The active neutrino mass matrix is defined in the framework of $\mathcal{O}(\theta^2)$ scenario by the seesaw type I equation
\begin{equation}
    m_\nu  \simeq  - M_D \theta^T \simeq  - M_D M^{-1}_M M_D^T
\end{equation}
with ambiguous definition of $M_D$ by means of $U_{PMNS}$ and $U_N$ mass matrix of the HNL sector
\hspace{-0.1cm}
\begin{equation}
\label{eq:md-omega-un}
    M_D = i U_{\rm PMNS} \sqrt{\hat{m}} \Omega \sqrt{\hat{M}} U_N^\dagger,
\end{equation}
where $\Omega$ is an arbitrary orthogonal matrix, $\Omega \Omega^T=I$. Components of the mixing matrix have the form
\hspace{-0.1cm}
\begin{equation}
\label{rv:general}
    	\Theta_{\alpha I} = i\frac{\sum_{k}\sqrt{m_k}~(U_\nu)_{\alpha k}\Omega_{k I}}{\sqrt{M_I}},
\end{equation}
where $\alpha = e,\mu,\tau$ and  $I=1,2,3$ is the number of HNL generation. Mixing for the first generation HNL, which is usually considered as a candidate for the role of a dark matter particle, is determined by the first column of the matrix $\Omega$.

In the simplest "minimal parametric mixing" case of the diagonal $\Omega=I$ matrix the mixing matrix for the normal mass hierarchy (NH) is
 \begin{eqnarray*}
 \label{eq:minmix:NH}
    \Theta^{\rm NH}_{\min}=\left( \begin{array}{ccc}
    iU_{e1}\sqrt{\frac{m_{1}}{M_{1}}} & iU_{e2}\sqrt{\frac{m_{2}}{M_{2}}} & iU_{e3}\sqrt{\frac{m_{3}}{M_{3}}} \\
    iU_{\mu1}\sqrt{\frac{m_{1}}{M_{1}}} & iU_{\mu2}\sqrt{\frac{m_{2}}{M_{2}}} & iU_{\mu3}\sqrt{\frac{m_{3}}{M_{3}}} \\
    iU_{\tau1}\sqrt{\frac{m_{1}}{M_{1}}} & iU_{\tau2}\sqrt{\frac{m_{2}}{M_{2}}} & iU_{\tau3}\sqrt{\frac{m_{3}}{M_{3}}}
    \end{array} \right)&&
\end{eqnarray*}
For inverted hierarchy (IH) the anti-diagonal $\Omega$ is used. One can observe using the HNL currents of Eq.(\ref{ncurr}) that the HNL production cross sections in the lowest order include the factor $|\Theta|^2$, so they are strongly suppressed by $m/M$ mass ratio.  
In the recent literature, more interesting choice is considered on an almost non-alternative basis
\vspace{-0.2cm}
 \begin{eqnarray*} 
    \Omega^{\rm NH}=\left(
    \begin{array}{ccc}
     1 & 0 & 0 \\
     0 & \cos (\omega ) & -\sin (\omega ) \\
     0 & \xi  \sin (\omega ) & \xi  \cos (\omega ) \\
    \end{array}
    \right) \\
    \text{or} \quad \Omega^{\rm IH}=\left(
    \begin{array}{ccc}
     0 & \cos (\omega ) & -\sin (\omega ) \\
     0 & \xi  \sin (\omega ) & \xi  \cos (\omega ) \\
     1 & 0 & 0 \\
    \end{array}
    \right)
\end{eqnarray*}
 Production enhancements appear with the complex-valued $\omega$ parameter which leads to the factors $X_{\omega}\,$=\,exp$(Im( \omega))$ in the mixing matrix $\Theta$. 
 An analysis of active and sterile neutrino mixing in \cite{asaka-eijima} demonstrated that a phenomenologically consistent hierarchy of mixings $\Theta_{e I}$, $\Theta_{\mu I}$ and $\Theta_{\tau I}$ with $\Theta_{e I}$ suppression relative to other mixing matrix elements can be achieved in a wide interval of $X_\omega$ independently on the values of HNL masses. The experimental upper bounds on $\Theta_{\alpha I}$ from the shortest possible lifetimes of $N_{2,3}$ from $\pi^\pm$ and $K^\pm$ meson decays converted into the upper bound on $X_\omega$ lead to $Im(\omega)=$4.5 at the HNL mass of the order of 10$^2$ MeV  for the lifetime of the order of 1 sec and $Im(\omega) \sim$7 for the lifetime of the order of 0.01 sec. Values of the $\omega$ parameter greater than seven can lead to a large mixing parameters not consistent with the EW data of high precision. Recent reconsideration in the light of modern data for HNL searches in the mass range less than the mass of 
$K$ meson has been performed in \cite{Bondarenko:2021cpc} for the case of two HNL generations. Model-dependence of $\Theta$-mixing results in a rather non-trivial set of consequences for the observables. For this reason collider studies are performed in the framework of the so-called "model-independent phenomenological approach"
\cite{alekhin} or, in other words, "phenomenological type I seesaw model" \cite{drewes}. 
In the model-independent phenomenological approach it is assumed that only a single HNL is available in an experiment, while other HNLs are sufficiently heavy and do not affect the analysis. There are only two independent parameters in this approach, the HNL mass and the Yukawa coupling defining HNL interaction with an active neutrino of a given flavor, assuming that the mixing with other flavors is zero. In the phenomenological type I seesaw model an additional parameter distinguishing Dirac neutrino case from Majorana neutrino case is added, which allows to discriminate observables specific for the field-theoretic model. Such simplifications are useful for derivation of generic bounds on the mixing parameter beyond any aspects of a particular model construction, but need an appropriate translation if one would like to go beyond the case of one generation and consider a well-defined mixing.

In the following analysis based on Casas-Ibarra diagonalization the mixing factor including $\sqrt{m/M}$ is enhanced by taking $Im(\omega)=3$, then evaluating $\Theta_{\alpha I}$. This approach is different from the specific mixing scenarios in the pseudo-Dirac limit, where an additional flavor symmetry is used to reconcile very small active neutrino masses with large mixings to give observable signals with HNL at the electroweak scale. Explicit expressions 
for the $\Theta_{\alpha I}$ as a function of $Im(\omega)$ are given in~\cite{asaka-eijima}. 

\vspace{-0.2cm}
\section{\label{sec:exp}Experimental procedure}

\vspace{-0.3cm}
\subsection{\label{sec:mc}Monte Carlo simulation and reconstruction}
\vspace{-0.1cm}

The signal processes $\ell^+\ell^- \to N W^{\pm} \ell^{\mp}$ with
$\ell = e, \mu$ are modelled using the CompHEP generator~\cite{comphep}, where the seesaw type-I model is
incorporated. 
As mentioned above, for simplicity only one Majorana neutrino is included in the calculations,
which is labeled as $N$ everywhere in the following text.
Other two HNLs decouple.
The calculations include the matrix elements for all possible diagrams at the parton level,
which are allowed for the studied process.
The 30 diagrams are obtained in the CompHEP generator to describe the signal process.
Two of these diagrams with $t$-channel vector bosons are shown in Fig.~1.

\vspace{-0.1cm}
\begin{figure}[H]\label{fig:mx} 
\centering
\vspace{-0.25cm}
\includegraphics[scale=0.17]{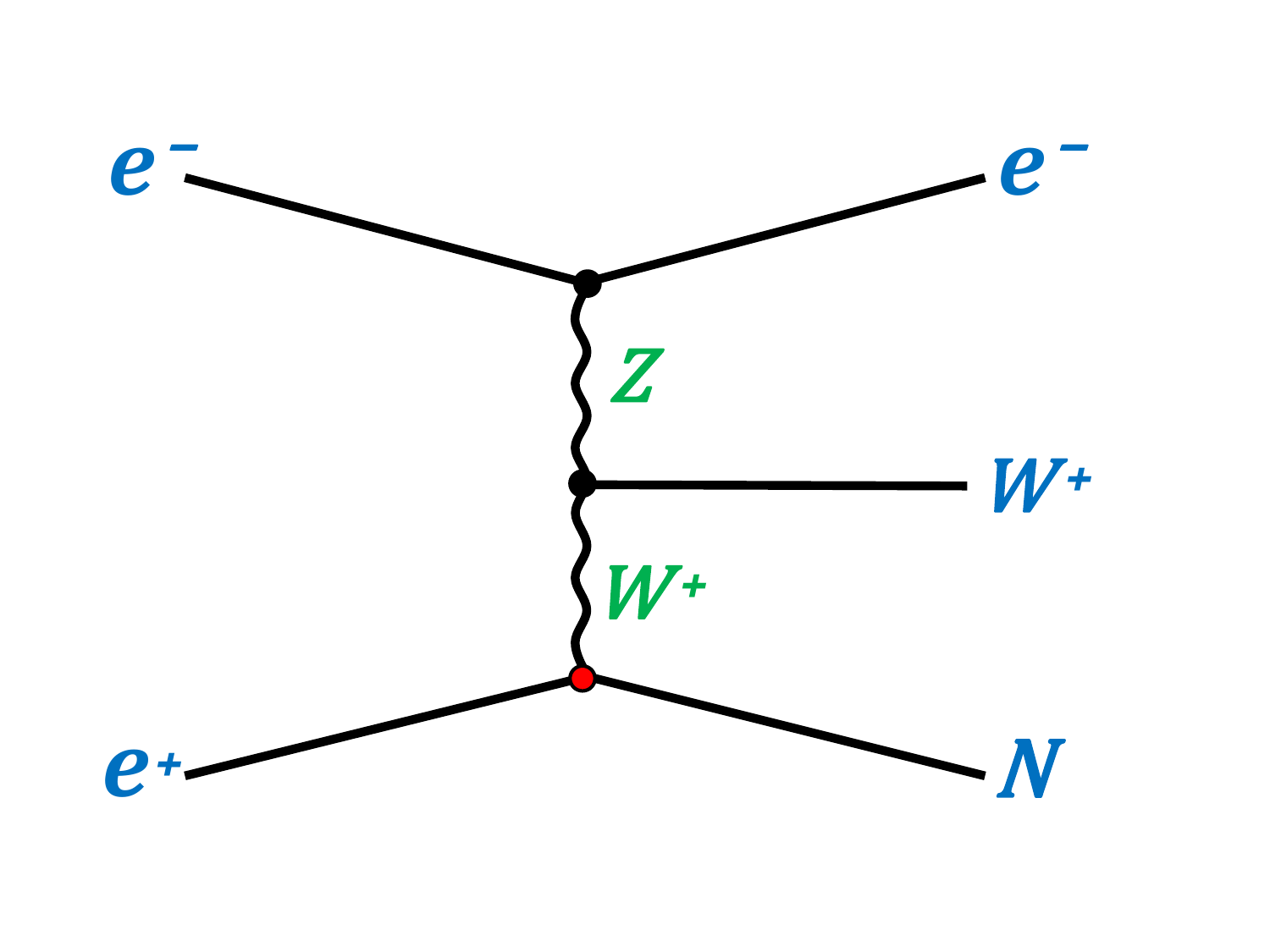}\hspace{-0.6cm}\includegraphics[scale=0.17]{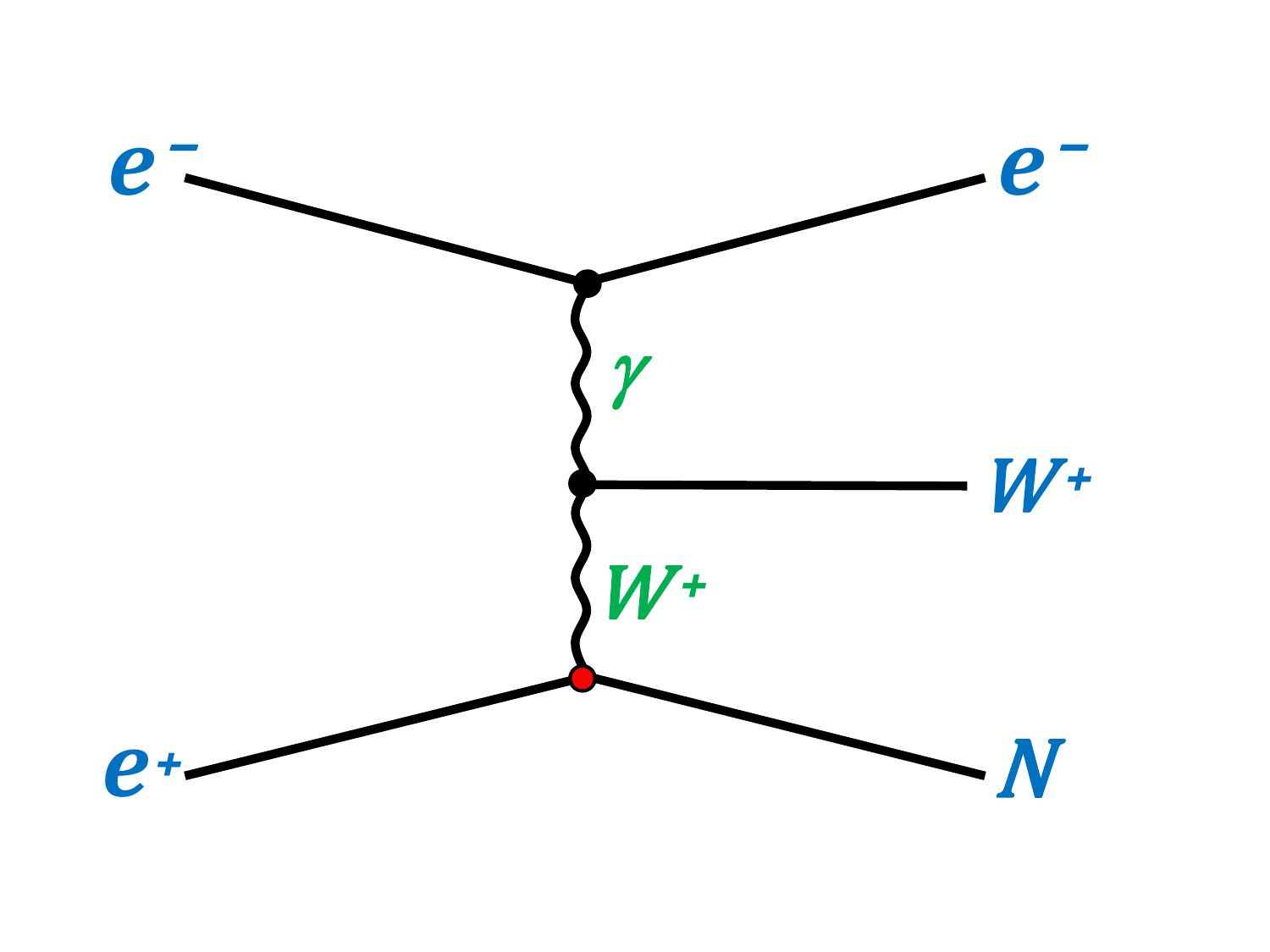}
\vspace{-0.45cm}
\caption{Two examples of diagrams describing the process $e^+e^-\to\,N W^+\,e^-$.}
\end{figure}

The CompHEP generator evaluates the cross sections for the studied processes and produces the Monte Carlo (MC) 
event data samples.
We generated the process $e^+e^- \to N W^+ e^-$ at the CM
energy 1~TeV and the process $\mu^+\mu^- \to N W^+ \mu^-$ at the CM energy 3 TeV and 10~TeV.
The beams are assumed to be unpolarized.
The HNL mass is varied in the range from 100~GeV to the CM energy.
The HNL width is assumed to increase with the HNL mass
similar to that obtained in~\cite{nne}.
The HNL width variation within 20\,\% results in cross section variation of less than 1\,\%.   

The subsequent decays $N \to W^{\pm} \ell^{\mp}$ and $W^{\pm} \to q\bar{q}$ are modelled by Pythia6.
Only hadronic decays of the $W$ bosons are used.
We forced the decay \mbox{angle $\theta_\ell^\star$} of the lepton in the HNL rest frame relative to the HNL direction 
to follow the \mbox{f($\theta_\ell^\star$) = (1 - cos($\theta_\ell^\star$))/2} distribution.
The ISR effects are included on the level of Pythia6.
Finally the quarks are hadronized by Pythia6. 
The SM background data samples are produced by the Whizard~2 generator which contains Pythia6.
The same beams and CM energies as for the signal samples are used to generate backgrounds.
The produced signal and background data samples
are processed with the Delphes program~\cite{delphes}, which
provides the fast and simplified detector simulation and the event reconstruction.
In Delphes the ILC detector card is used in case of the $e^+e^- \to N W^+ e^-$ process at 1~TeV,
and the MuC detector card is used for the $\mu^+\mu^- \to N W^+ \mu^-$ process at 3 TeV and 10 TeV.
The jets are reconstructed using the Valencia algorithm with the default parameters.
The Delphes jet algorithm is forced to reconstruct exactly four jets. 
The produced output data samples comprise the information about the four-momenta of the reconstructed jets
and the isolated leptons. The signal and background data samples produced by Delphes are analyzed
using the ROOT package.

\vspace{-0.3cm}
\subsection{\label{sec:mc}Cross sections}
\vspace{-0.1cm}

The cross sections for the $\ell^+\ell^- \to N W^+ \ell^-$ processes calculated by the CompHEP program are shown in Fig.~2.
The mixing parameter is fixed to $|V_{\ell N}|^2 = 0.0003$ to compare our results with~\cite{nne},
where this value was used.
The cross sections for the $\ell^+\ell^- \to N \nu_{\ell}$
processes computed by \mbox{CompHEP} are also shown in Fig.~2. The same beams, CM energies,
and mixing parameters are assumed for both processes. Unitary cancellation of the second order $t$-channel pole, see Fig. 1 (right diagram) is checked directly using $d\sigma/d({\rm ln}\ t)$ distribution \cite{boos}.
The $\ell^+\ell^- \to N \nu_{\ell}$ process cross sections obtained with CompHEP agree perfectly
with the cross sections calculated in~\cite{nne} using the Whizard 2 generator.

\begin{figure}\label{fig:mx} 
\centering
\hspace{-0.55cm}\includegraphics[scale=0.45]{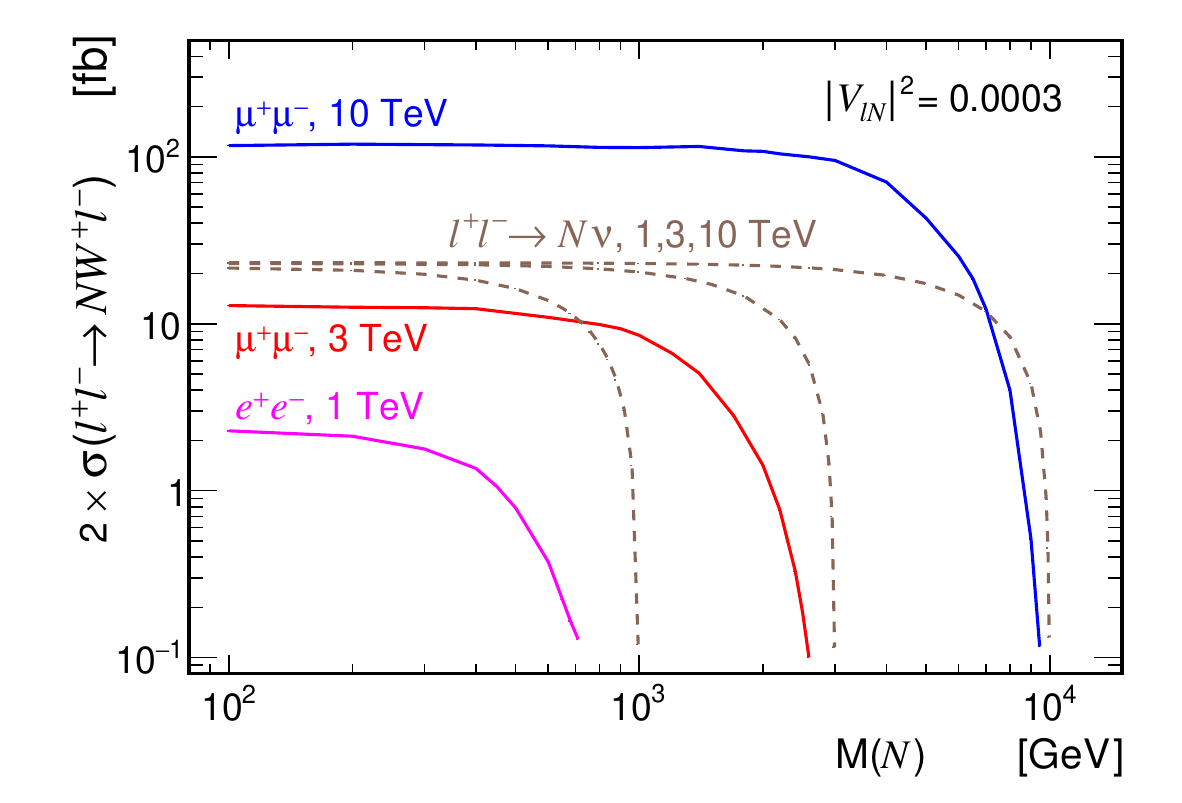}
\vspace{-0.1cm}
\caption{The cross sections as a function of the HNL mass are shown for the process
$e^+e^- \to N W^+ e^-$ at 1~TeV (purple line) and the process $\mu^+\mu^- \to N W^+ \mu^-$
at 3 TeV (red line) and 10 TeV (blue line). The factor 2 reflects two charge-conjugated final states.
For comparison the cross sections for the processes $\ell^+\ell^- \to N \nu_{\ell}$
for the same mixing parameter, beam particles and CM energies are also shown (dashed brown lines).}
\end{figure}

As we can see in Fig.~2, the plateau in the cross section of the $\ell^+\ell^- \to N W^{\pm} \ell^{\mp}$ process
is growing with the CM energy in contrast to the $\ell^+\ell^- \to N \nu_{\ell}$ process. 
At high CM energies the process $\ell^+\ell^- \to N W^{\pm} \ell^{\mp}$ has large cross sections and can be used 
to obtain strong upper limits on the mixing parameters. Additional advantages of this process are 
a high signal reconstruction efficiency and a high signal to background separation ratio.
\vspace{-0.3cm}
\subsection{\label{sec:mc}Selections}
\vspace{-0.1cm}

The process $\ell^+\ell^- \to W^+\,W^+\,\ell^-\,\ell^-$ includes two same-sign leptons and four jets in the final state.
Although there is no SM background to this final state, the reconstruction efficiency is 
below a percent level. 
One of the final state leptons is mostly emitted close to the beam direction and
is not registered in detectors or badly measured in forward calorimeters.
Therefore we will not discuss the signature with all particles reconstructed.

We select the final states with four jets and one lepton.
The initial preselections are applied on the energy and pseudorapidity of the jets and the lepton:

\begin{fleqn}[\parindent]
\begin{equation}\label{eq:presel}
\begin{split}
& E(j) > 10~{\rm GeV};   \ \ \  E(\ell) > 10~{\rm GeV};  \\
&|\eta(j)| < 2.5;  \hspace{0.96cm} |\eta(\ell)| < 2.5
\end{split}
\end{equation}
\end{fleqn}

The jets are produced in the $W$ boson decays, respectively we combine
the four jets in the two bosons. Among possible combinations, the two jet pairs with the invariant masses closest
to the nominal $W$ boson mass are chosen as the $W$ candidates. If a $W$ boson has a large energy,
the produced jets can be partially overlapped. In this case the parameters of the individual jets can be incorrect,
however the respective two-jet combination reproduces the $W$ boson mass and momentum with a reasonable accuracy.

To separate signal events from backgrounds we use the following variables:
\renewcommand{\labelenumi}{\alph{enumi})}
\begin{enumerate}
\setlength\itemsep{0.1em}
\item $M_{miss} (4j\,\ell^-)$ - the missing mass to the sum of the four jets and the reconstructed lepton \vspace{0.1cm}
\item $M(jj)$ - the masses of jet pairs, corresponding to the $W$ boson candidates
\item cos$(W_1 \vee W_2)$ - cosine of the angle between two $W$ candidates
\item cos$(\ell^- \vee \ell^-_{beam})$ - cosine of the angle between the directions of the reconstructed lepton and the same-sign beam lepton
\item cos($P_{miss} (4j\,\ell^-) \vee \ell^-_{beam}$) - cosine of the angle between the missing momentum (to
the four-jet and the isolated lepton) and the same-sign beam lepton
\item $M_{\Delta}(W\ell^-)$ - the invariant masses of the $W$ boson and the reconstructed lepton (two combinations)
\end{enumerate}

The cut on the variable $M_{miss} (4j\,\ell^-)$ is used to remove the backgrounds with two or more unobserved particles,
because such backgrounds must have a large missing mass.
The mass of the jet pairs is required to lie in the range $50 < M(jj) < 120\,$GeV for the selected
$W$ candidates.
The backgrounds with only one produced $W$ or $Z$ boson decaying in two jets can 
imitate four-jet configuration. This background must have a small angle between the $W$ candidates,
and the cut cos$(W_1 \vee W_2) < 0.8$ is applied to suppress this background.
The backgrounds coming from the vector boson fusion processes should result in a small lepton angle
relative to the corresponding beam direction. The cut cos$(\ell^- \vee \ell^-_{beam}) < 0.7$ is used to
provide a significant suppression of these backgrounds. This cut is also suppresses the backgrounds,
where a lepton is produced in the decay of the $W$ boson.
As it was explained above, the unobserved lepton is mostly emitted with a small angle relative to the direction of the 
corresponding beam. The sign of the beam is fixed by the sign of the reconstructed lepton.
Therefore the momentum of the system of the four jets and lepton can be used to estimate the direction 
of the missed lepton. The variable cos($P_{miss} \vee \ell^-_{beam}$) is
peaked at 1 for the signal events, except the events with the NHL mass close to E$_{\rm cms}$.
In contrast, the backgrounds due to vector boson fusion production peak at -1,
whereas other backgrounds have flat distributions.
Therefore the cut cos($P_{miss} \vee \ell^-_{beam}) > 0.9$ is applied for all HNL mass values,
except the high HNL mass region, where this signal distribution becomes flat.
The applied cuts are summarized in Table I.
Finally, the distributions of the invariant mass of the $W$ boson and the reconstructed lepton 
$M_{\Delta}(W\ell^-$) have to be searched for a peak corresponding to the HNL production.
The mass difference $M_{\Delta}(W\ell^-) =  M(jj\ell^-) - M(jj$) + 80.377 is used instead of $M(jj\ell^-$)
to improve the signal resolution especially at the low HNL mass region up to 500~GeV.

\vspace{-0.3cm}
\renewcommand{\arraystretch}{1.2}
\begin{table}[htb]
\caption{The cuts on variables applied at different CM energies.
The last column shows the cuts applied at the high HNL mass region M($N) >$ 0.85 E$_{\rm cms}$,
where the cut on cos($P_{miss} \vee \ell^-_{beam}$) is removed.}
\vspace{-0.02cm}
\begin{center}
\label{tab2}
\begin{tabular}
{@{\hspace{0.1cm}}l@{\hspace{0.2cm}} @{\hspace{0.2cm}}c@{\hspace{0.2cm}} @{\hspace{0.2cm}}c@{\hspace{0.2cm}} @{\hspace{0.2cm}}c@{\hspace{0.2cm}} @{\hspace{0.2cm}}c@{\hspace{0.1cm}}}
\hline \hline
CM energy, & 1 TeV & 3 TeV & 10 TeV & M($N) >$  \\
beams           & $e^+e^-$ & $\mu^+\mu^-$ & $\mu^+\mu^-$ & 0.8 E$_{\rm cms}$ \\
\hline
 $M_{miss}$, GeV & $< 350$ & $< 800$ & $< 3000$ & same \\
 $M(jj)$, GeV & 50-120 &  50-120 &  50-120 &  50-120 \\
 cos$(W_1 \vee W_2)$ & $< 0.8$ & $< 0.8$ & $< 0.8$ & $< 0.8$ \\
 cos$(\ell^- \vee \ell^-_{beam})$ & $< 0.7$ & $< 0.7$ & $< 0.7$ & $< 0.7$ \\
 cos($P_{miss} \vee \ell^-_{beam}$) & $> 0.9$ & $> 0.9$ & $> 0.9$ & - \\
\hline \hline
\end{tabular}
\end{center}
\end{table}
\vspace{-0.2cm}
\subsection{\label{sec:mc}Backgrounds}

For simplicity, the electron beam channel is considered in this section,
however almost the same analysis is performed for the muon beam channels.
The studied backgrounds are required to have one identified isolated lepton.
The isolated leptons can be produced in the $W$ or $Z$ boson decays or in the vector boson fusion processes.
Backgrounds with a faked lepton are not studied here, these backgrounds are expected to give a small contribution.
The four jets are required to be reconstructed, however there are backgrounds where the two jets are splitted
and imitate the four jets. 

We assumed an integrated luminosity of 1 ab$^{-1}$ for both $e^+e^-$ collisions at 1 TeV and $\mu^+\mu^-$
collisions at 3~TeV. An integrated luminosity of 10 ab$^{-1}$ is assumed for the $\mu^+\mu^-$
collisions at 10 TeV.
The cross sections for all potentially dangerous backgrounds are evaluated using the Whizard 2 generator,
and the approximate numbers of events expected to contribute in the $M_{\Delta}(W\ell^-)$ distribution are calculated.
If the estimated numbers of events are not too small, the event data samples are generated to obtain
a more accurate estimates. 
The pseudorapidity cut $|\eta(\ell^-)| < 2.6$ is applied on the level of the background generation for the negative
leptons directly produced in the vector boson fusion processes.
The 6 background channels are generated, which are supposed to be potentially dangerous:

\begin{enumerate}
\setlength\itemsep{-0.1em}
\item  \ \ $e^+e^- \to W^+(q\bar{q})\,e^- \nu_{e}$
\item  \ \ $e^+e^- \to W^+(q\bar{q}) W^-(e^-\nu_{e})$
\item  \ \ $e^+e^- \to W^+(q\bar{q}) Z(q\bar{q})\,e^- \nu_{e}$
\item  \ \ $e^+e^- \to W^+(q\bar{q}) W^-(q\bar{q})\,e^+e^-$
\item  \ \ $e^+e^- \to  W^+(q\bar{q}) Z(q\bar{q}) W^-(e^-\nu_{e})$
\item  \ \ $\gamma^B e^- \to W^+(q\bar{q}) W^-(q\bar{q})\,e^-$
\end{enumerate}

\begin{figure*}[t!]\label{fig:var} 
\centering
\includegraphics[scale=0.303]{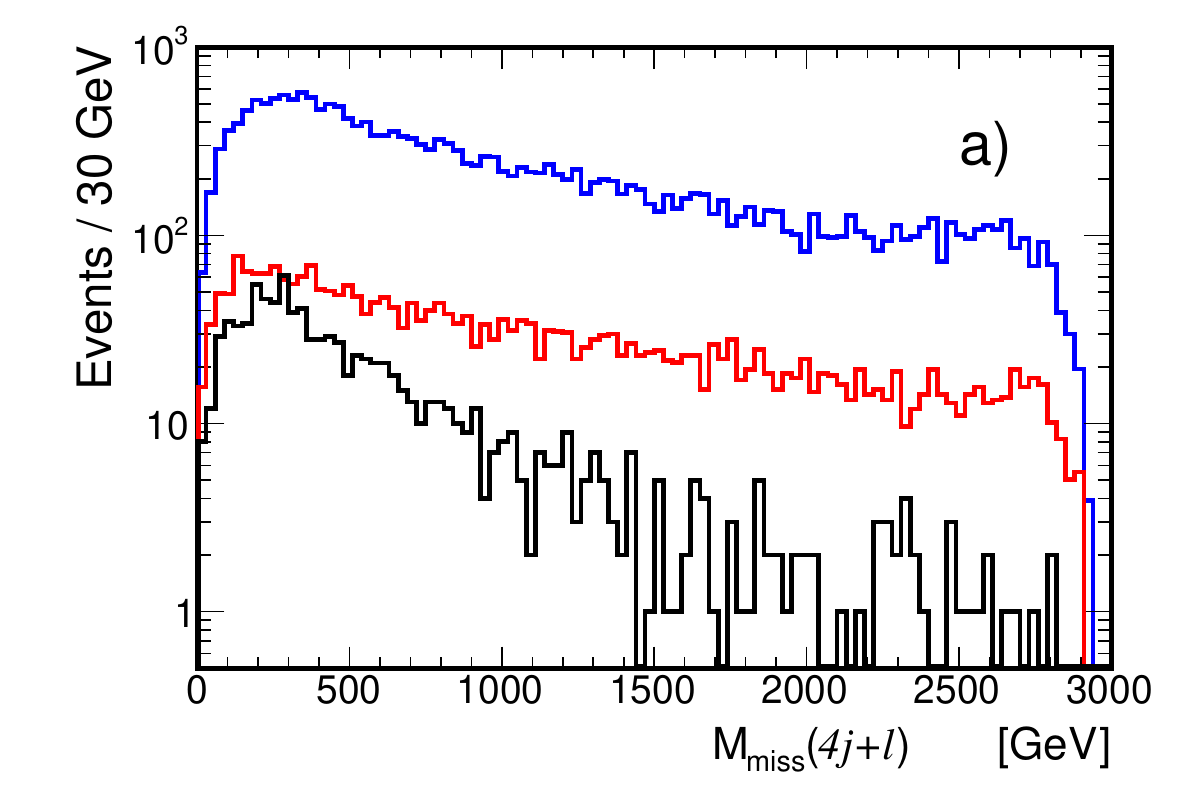}\includegraphics[scale=0.303]{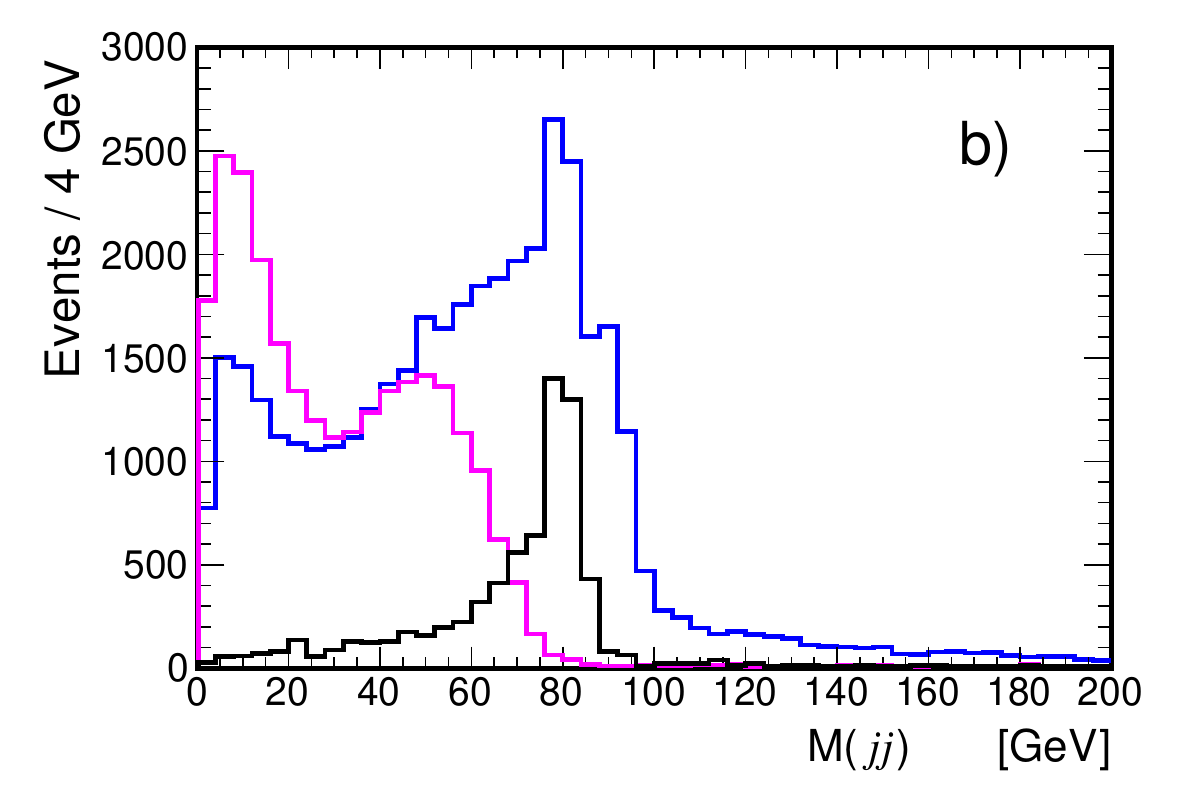}\includegraphics[scale=0.303]{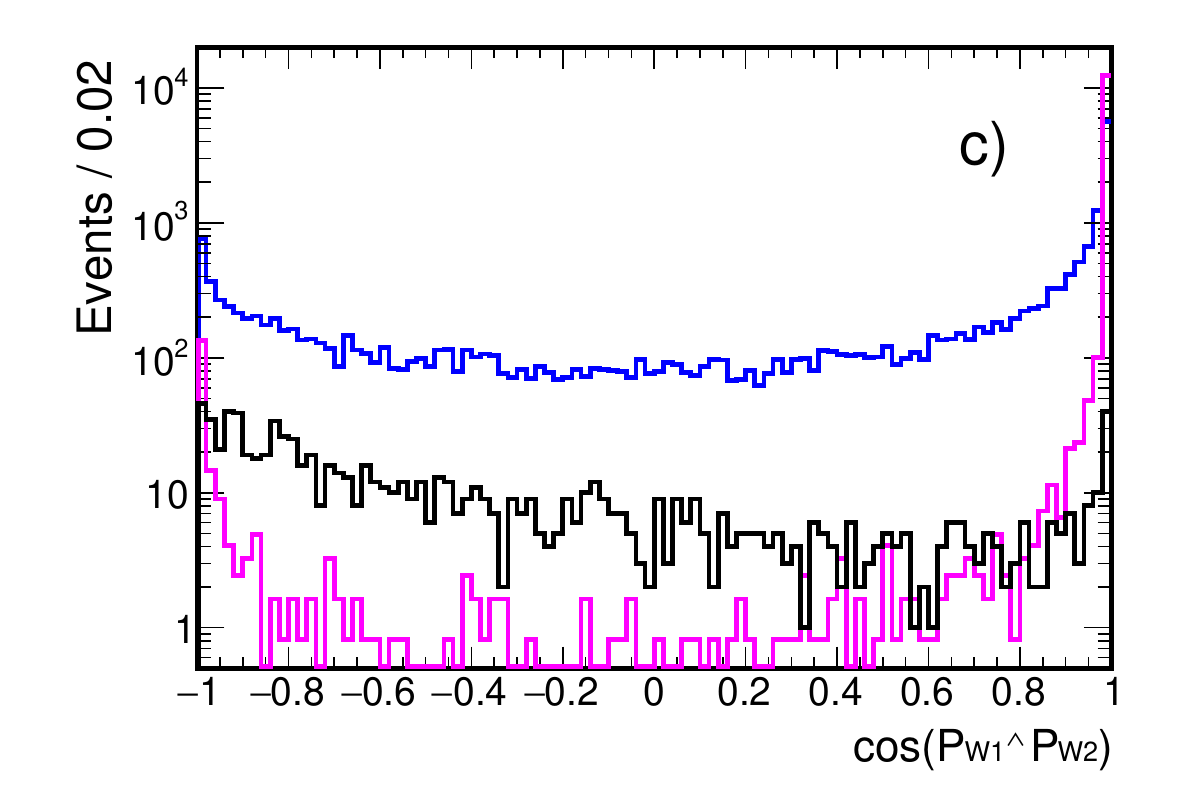}
\includegraphics[scale=0.303]{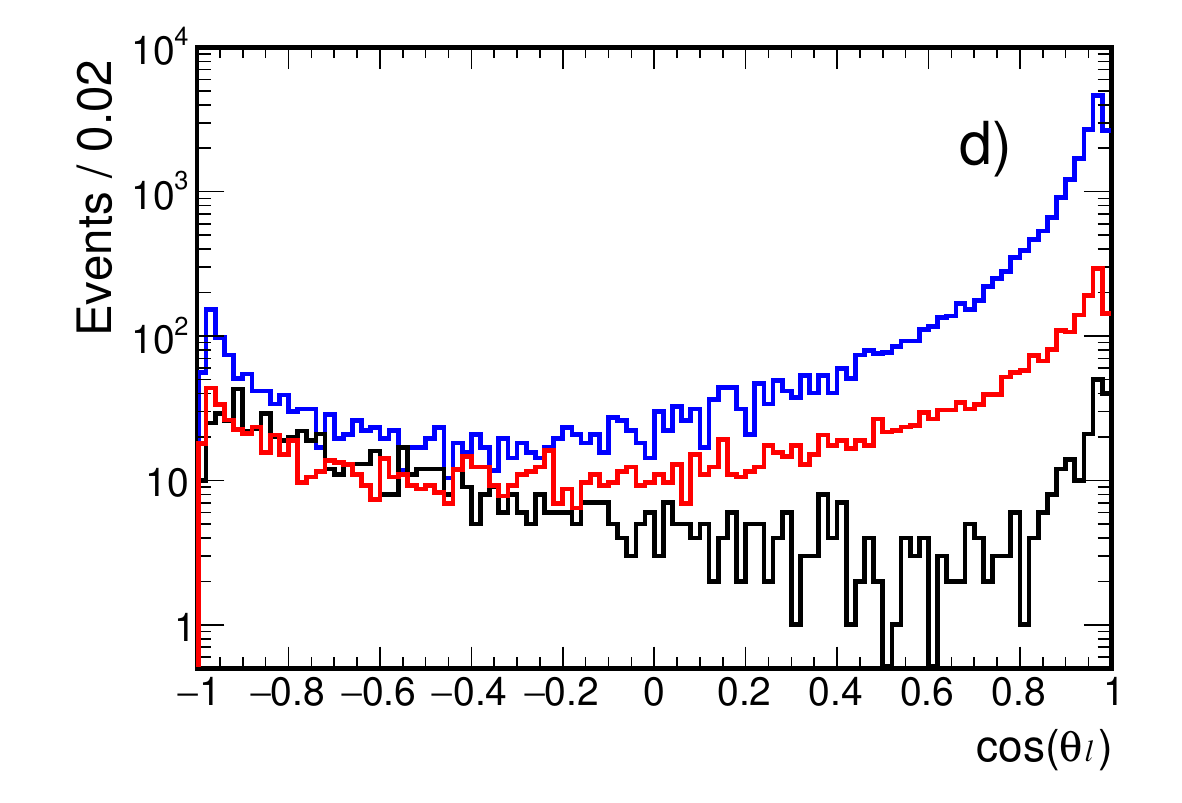}\includegraphics[scale=0.303]{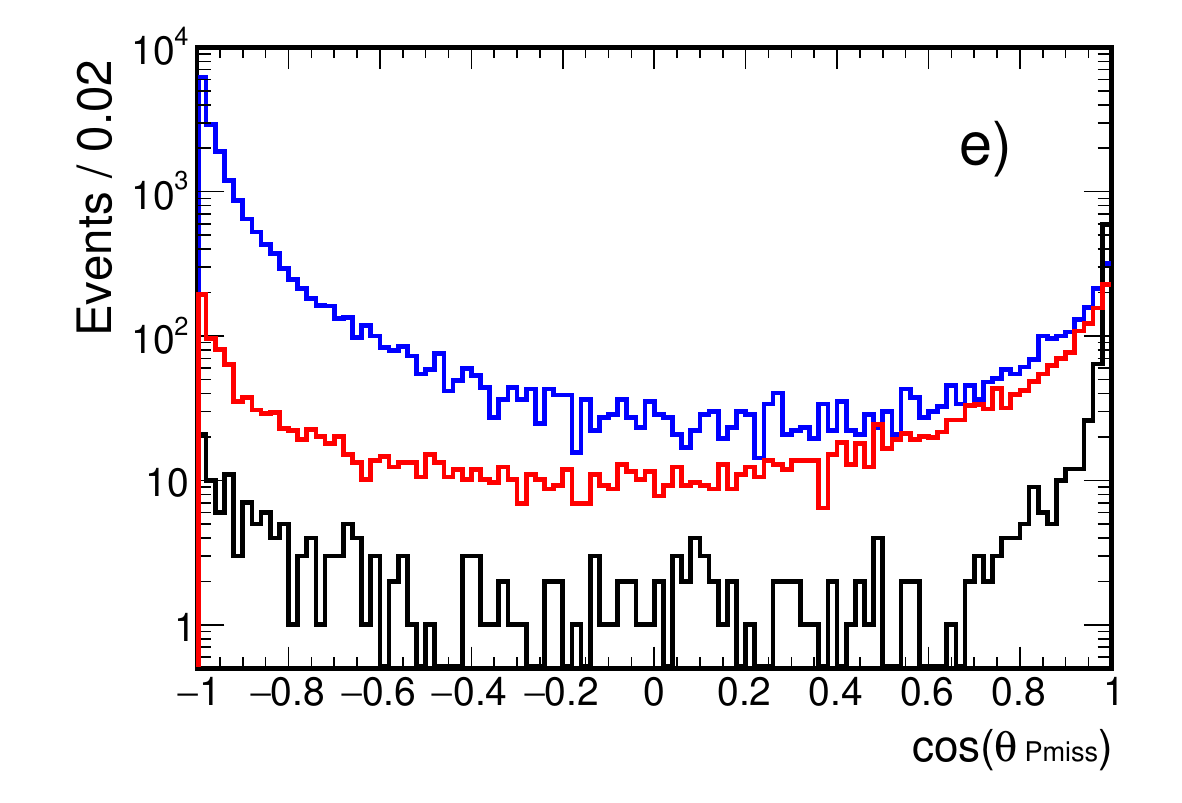}\includegraphics[scale=0.303]{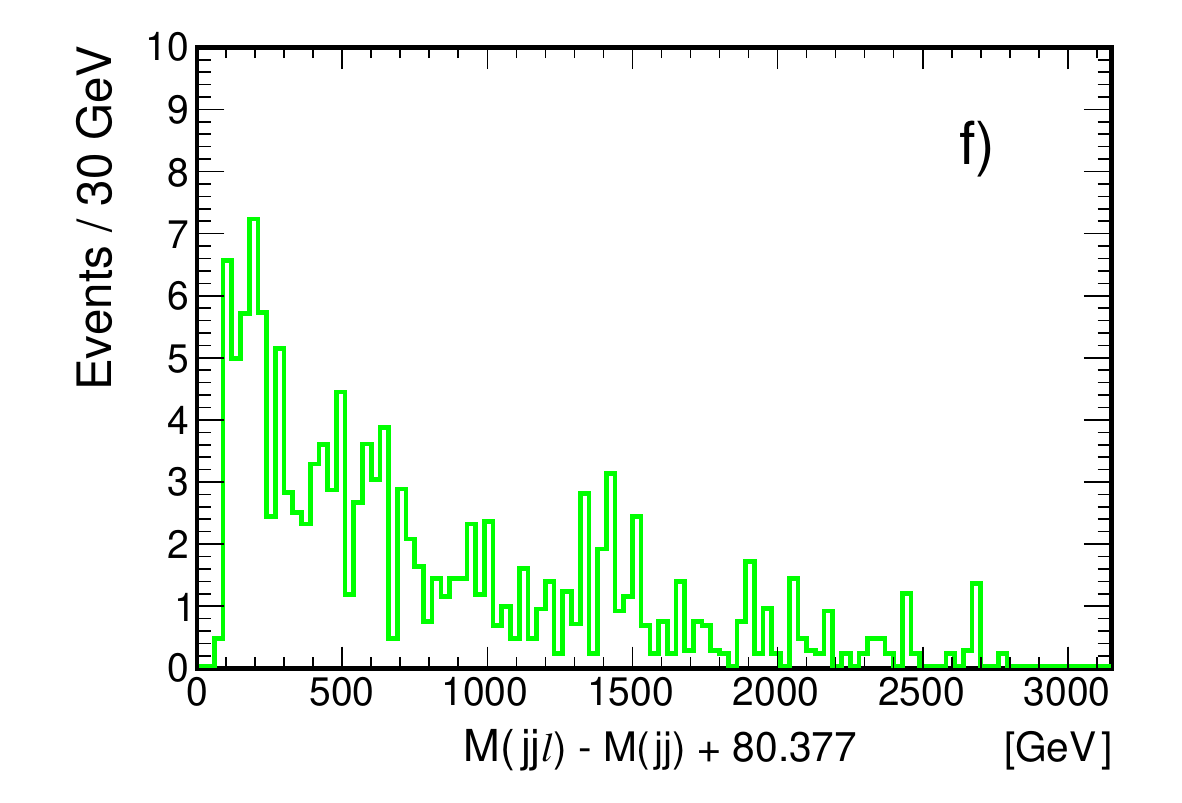}
\caption{The distributions for the a) $M_{miss} (4j\,\ell^-)$, b) $M(jj)$, c) cos$(W_1\,\vee\,W_2)$,
d) cos$(\ell^-\,\vee\,\ell^-_{beam})$, e) \mbox{cos($P_{miss} (4j\,\ell^-) \vee \ell^-_{beam}$)} and f) $M_{\Delta}(W\ell^-)$
parameters are shown for the $\mu^+\mu^-$ collisions at CM energy of 3 TeV. 
The black histograms are obtained for the signal process $\mu^+\mu^- \to N W^+ \mu^-$ with M($N$) = 600 GeV.
The signal in figure b) is multiplied by factor 4. 
The color histograms show the background distributions for the processes $\mu^+\mu^- \to W^+(q\bar{q}) Z(q\bar{q})\,\mu^- \nu_{\mu}$ (blue),
$\mu^+\mu^- \to  W^+(q\bar{q}) Z(q\bar{q}) W^-(\mu^-\nu_{\mu})$ (red) and $\mu^+\mu^- \to W^+(q\bar{q}) W^-(\mu^-\nu_{\mu})$ (purple).
The distributions are obtained after preselections, but before main cuts. The background distributions are normalized to 
the cross sections, the signal distributions are normalized to 1000 events.
The green histogram in figure f) shows the sum of all background distributions at 3 TeV after all cuts.}
\end{figure*}

Here the photons produced from beamstrahlung process are assigned as $\gamma^B$.
Figure 3 demonstrates the distributions of the introduced above variables for specific backgrounds at 3 TeV
$\mu^+\mu^-$ collisions.
The corresponding distributions at 1 TeV and 10 TeV have very similar shapes. 

As we can see from the Fig. 3c, the backgrounds with only one $W$ boson
are strongly suppressed by the cut cos$(W_1 \vee W_2) < 0.8$.
The backgrounds $e^+e^- \to q\bar{q}\,e^- \nu_{e}$ with mass M($q\bar{q}$) $>$ 120~GeV,
which are not associated with the $W$ boson, have cross sections of a few fb before cuts,
that results in contributions of about 1 event or less.
The cross section of the $e^+e^- \to Z(jj) Z(jj)\,e^+ e^-$ process is about (1-3) fb at
the studied CM energies and the contributions from this process are less than 2 events.
The backgrounds $\gamma^B \mu^- \to W^+(q\bar{q}) W^-(q\bar{q})\,\mu^-$ give
a very small contribution because of the large muon mass and the round beams. 
The backgrounds due to the $\gamma\gamma$ collisions are suppressed by the $M_{miss}$ cuts.
The background from the $e^+e^- \to t\bar{t} \to W^+W^-b\bar{b}$ process is strongly suppressed
by the $M(jj) <$ 120~GeV cut.

The production cross sections obtained from Whizard and the numbers of events
in the background channels obtained from the generated samples
after all cuts are given in Table II.
We obtained the marginal contributions from the backgrounds a) and b) of about a few events
after all cuts. Therefore the background channels with only one vector boson can be neglected.
The cross section for the $\gamma e^- \to W^+(q\bar{q}) W^-(q\bar{q})\,e^-$ background
depends on beam parameters and is estimated with a large uncertainty. However the number of event
in this channel is small and this contribution can also be neglected. 
Although the suppression factors for the background channels c), d), e) are relatively moderate,
the cross sections of these backgrounds are not too large. Respectively, 
the $M{_\Delta}(W\ell^-)$ distribution with all backgrounds comprised has
a relatively small number of events (Fig.~3f).
Both combinations of the $W$ boson and the lepton are included in the $M{_\Delta}(W\ell^-)$ distribution.

\renewcommand{\arraystretch}{1.2}
\begin{table}[htb]
\caption{The cross sections in femtobarn for the studied background channels 
and the estimated numbers of events after all cuts (in brackets). The cut $|\eta(\ell^-)| < 2.6$ is applied
on the generation level for the directly produced negative leptons.}
\vspace{0.01cm}
\begin{center}
\label{tab2}
\begin{tabular}
{@{\hspace{0.1cm}}l@{\hspace{0.2cm}} @{\hspace{0.2cm}}c@{\hspace{0.1cm}} @{\hspace{0.2cm}}c@{\hspace{0.1cm}} @{\hspace{0.1cm}}c@{\hspace{0.1cm}}}
\hline \hline
CM energy, beams & 1 TeV & 3 TeV & 10 TeV  \\
\cline{1-1}
Final state          & $e^+e^-$ & $\mu^+\mu^-$ & $\mu^+\mu^-$ \\
\hline
$W^+(q\bar{q})\,\ell^- \nu_{\ell}$ &     939 (1)  &   575 (2)  & 151 (3) \\                 
$ W^+(q\bar{q}) W^-(\ell^-\nu_{\ell})$  & 3042 (1) & 538 (1) &  73.1 (0) \\
$W^+(q\bar{q}) Z(q\bar{q})\,\ell^- \nu_{\ell}$ &  11.2 (66) & 25.3 (26) & 22.4 (73) \\
$W^+(q\bar{q}) W^-(q\bar{q})\,\ell^+\ell^-$ &  29.6 (54) & 69.5 (9) & 51.7 (36) \\
$W^+(q\bar{q}) Z(q\bar{q}) W^-(\ell^-\nu_{\ell})$ &  57.7 (84) & 34.5 (32) & 10.1 (61) \\
$\gamma e^-\hspace{-0.1cm}\to\hspace{-0.03cm}W^+(q\bar{q}) W^-(q\bar{q})\,e^-$ & 40.2 (1)& - & - \\
\hline \hline
\end{tabular}
\end{center}
\end{table}

\vspace{-0.7cm}
\subsection{\label{sec:mc}Upper limits on mixing parameters}

The signal events are generated by CompHEP for three CM energies and different HNL masses.
Each data sample contains 1000 events, where only $W \to q\bar{q}$ decays are permitted.
The reconstructed signals after all cuts are shown in Fig.~4.
The CM energies, the HNL masses and the numbers of the signal events in the corresponding mass
windows are given in Table III. 

\begin{figure*}[t!]\label{fig:sig} 
\centering
\includegraphics[scale=0.303]{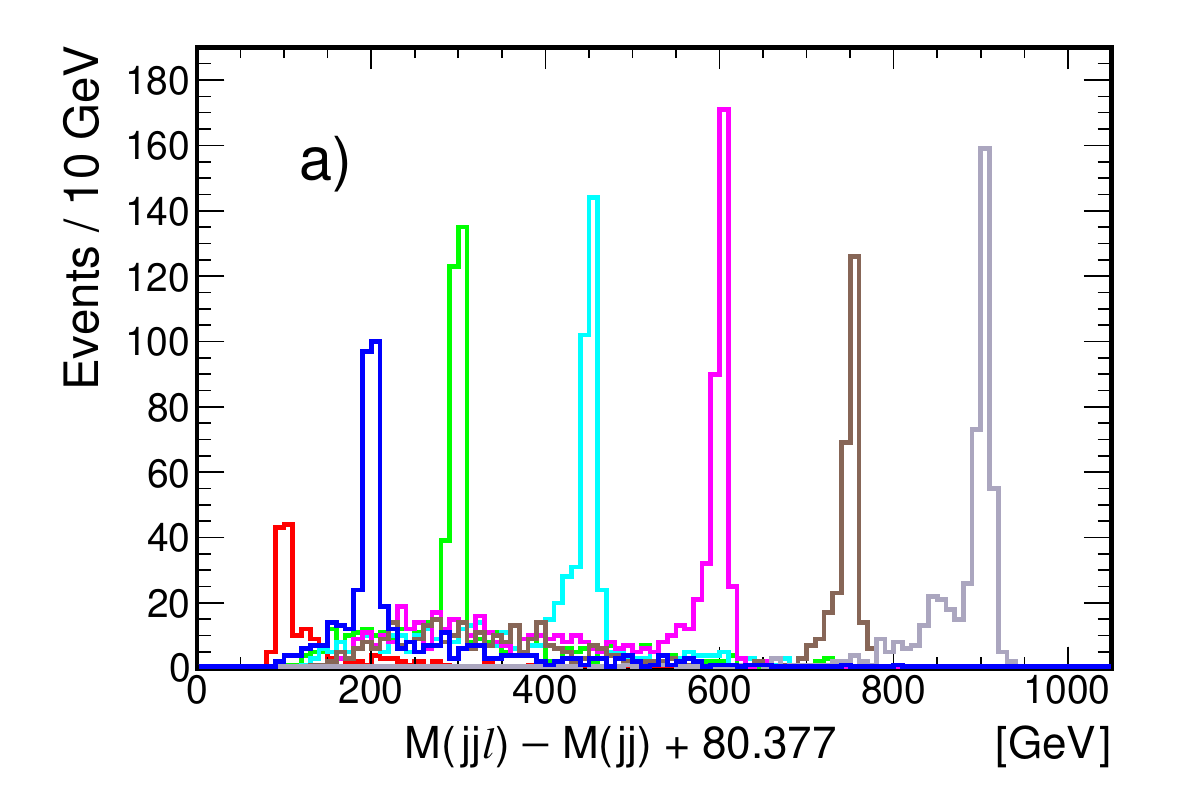}\includegraphics[scale=0.303]{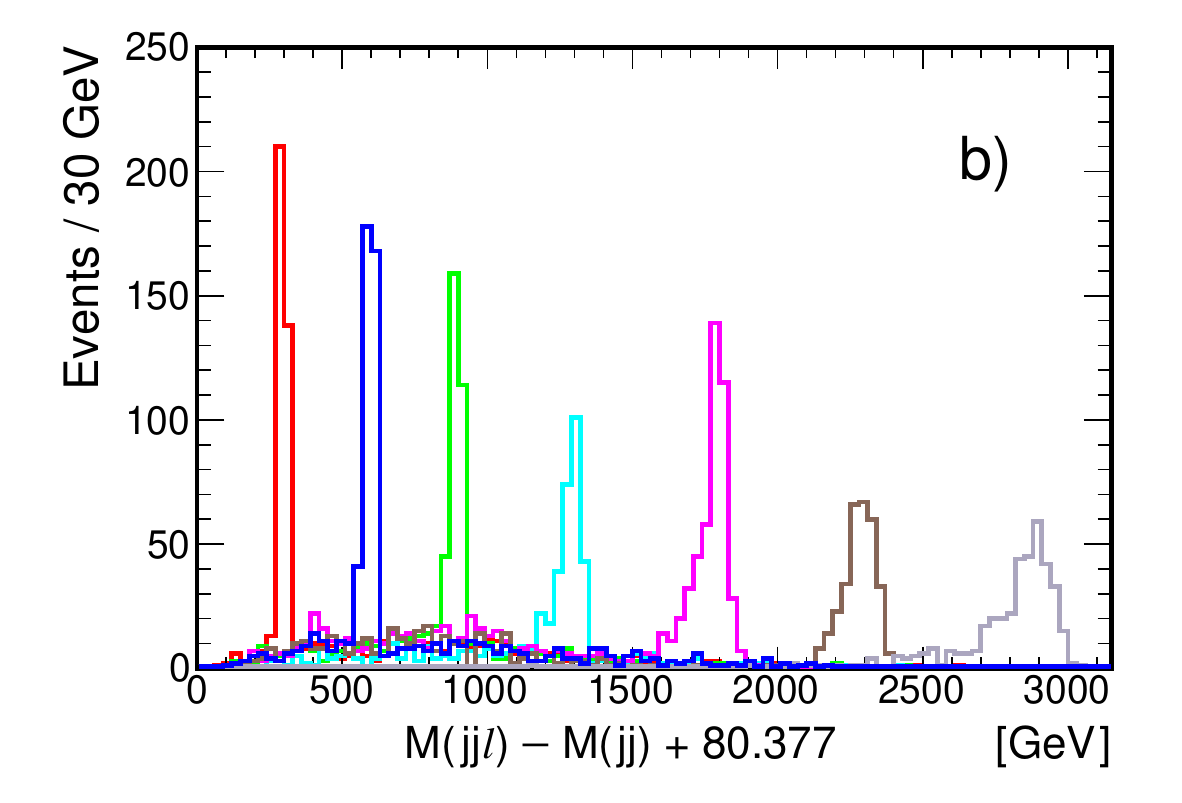}\includegraphics[scale=0.303]{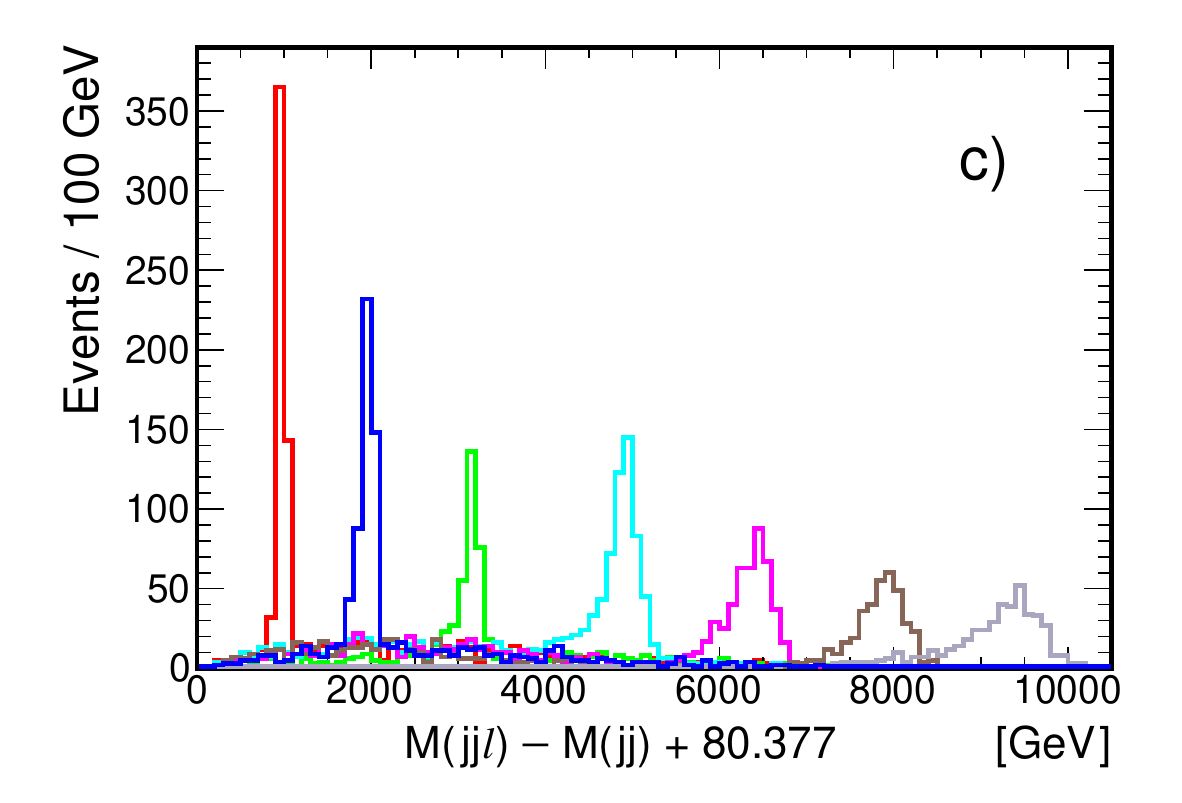}
\caption{The signal $M_{\Delta}(W\ell^-)$ mass distributions are shown for a) the $e^+e^-$ collisions at 1 TeV,
b) $\mu^+\mu^-$ collisions at 3 TeV, and c) $\mu^+\mu^-$ collisions at 10 TeV. The signals for the different HNL
masses are shown in different colors. Two combinations of the $W$ boson and the lepton are included,
except the HNL with the largest mass, where only the combination with the high momentum $W$ is taken.
The values of the modelled HNL masses are given in Table III.}
\end{figure*}

\renewcommand{\arraystretch}{1.2}
\begin{table}[htb]
\caption{The modelled HNL masses, the corresponding production cross sections, and the numbers of signal events in the 
chosen mass windows are given for the studied CM energies and beams. The data samples of 1000 events are generated for each HNL mass value,
only $W \to q\bar{q}$ decays are permitted.}
\vspace{0.01cm}
\begin{center}
\label{tab2}
\begin{tabular}{lcccccccc}
\hline \hline
\multicolumn{9}{c}{1 TeV,  $e^+e^-$} \\ \hline
$M(N)$, GeV & 100 & 200 & 300 & 450 & 600 & 750 & 900 & \\
$\sigma$, fb & 2.28 &  2.12 &  1.78 & 1.06 & 0.31 & 0.11 & 0.002 & \\
N events & 79 & 208 & 268 & 251 & 268 & 203 & 283 & \\ \hline
\multicolumn{9}{c}{3 TeV,  $\mu^+\mu^-$} \\ \hline
$M(N)$, GeV & 300 & 600 & 900 & 1300 & 1800 & 2300 & 2900 & \\
$\sigma$, fb & 12.4 & 10.9 & 9.33 & 5.85 & 2.33 & 0.53 & 0.01 & \\
N events & 322 & 358 & 303 & 254 & 400 & 295 & 297 & \\ \hline
\multicolumn{9}{c}{10 TeV,  $\mu^+\mu^-$} \\ \hline
$M(N)$, GeV & 300 & 1000 & 2000 & 3200 & 5000 & 6500 & 8000 & 9500 \\
$\sigma$, fb &  118.4 & 113.5 & 107.8 & 90.1 & 42.8 & 18.6 & 3.99 & 0.1 \\
N events & 349 & 462 & 426 & 263 & 457 & 356 & 291 & 294 \\
\hline \hline
\end{tabular}
\end{center}
\end{table}

To obtain the numbers of signal events the mass windows are chosen respectively to the width of the signals.
At the electron beam channels the mass windows are $\pm$10~GeV, except the first and last mass values.
At the muon beam channels the mass window is $[0.94 - 1.03]\,\times\,M(N)$, except the first and last
mass values. The windows for the first and last mass values are specially adjusted to observed widths of the signals.
To estimate backgrounds under the signals the numbers of background events are obtained in the same
mass windows. 

We tested the signal reconstruction efficiencies for various angular distributions in the $N \to W^{\pm} \ell^{\mp}$ decay.
Signal events are generated assuming uniform and different linear cos($\theta_\ell^\star$) distributions.
The efficiencies obtained after all cuts are approximately the same for all options within the $\pm\,5\,\%$ interval.
This follows from the fact that the acceptances and the reconstruction efficiencies of the
lepton and $W$ boson produced in the decay are similar.

Finally, the upper limits on the mixing parameters $|V_{\ell N}|^2$ as a function of $M(N)$ are
calculated (Fig.~5). The upper limits are obtained assuming $2\sigma$ signal under the background level but
not less than the 8 signal events. The decay branching fractions $Br(N \to W^+\ell^-)$ are not known and
we normalized the obtained upper limits to this value. It has to be noted, that 
the HNL can potentially decay in the modes $N \to W^+e^-$ and $N \to W^+\mu^-$.
If the both modes are used in analysis, an additional suppression of the boson fusion induced 
background is expected, however this effect is small.

As we can see in Fig.~5, the upper limits obtained in the process $e^+e^- \to N W^{\pm} e^{\mp}$
at 1 TeV are somewhat worse than the ones obtained in the process $e^+e^- \to N \nu_{e}$ \cite{nne}.
However, the upper limits for the process $\mu^+\mu^- \to N W^{\pm} \mu^{\mp}$ are competitive at 3 TeV and
overtake the process $\mu^+\mu^- \to N \nu_{\mu}$ \cite{Felix} at 10~TeV
\cite{Felix, kwok, li}. In these publications the current and future LHC upper limits on the mixing parameters are also 
shown.
\vspace{0.5cm}

\begin{figure}[h!]\label{fig:sensi} 
\centering
\includegraphics[scale=0.45]{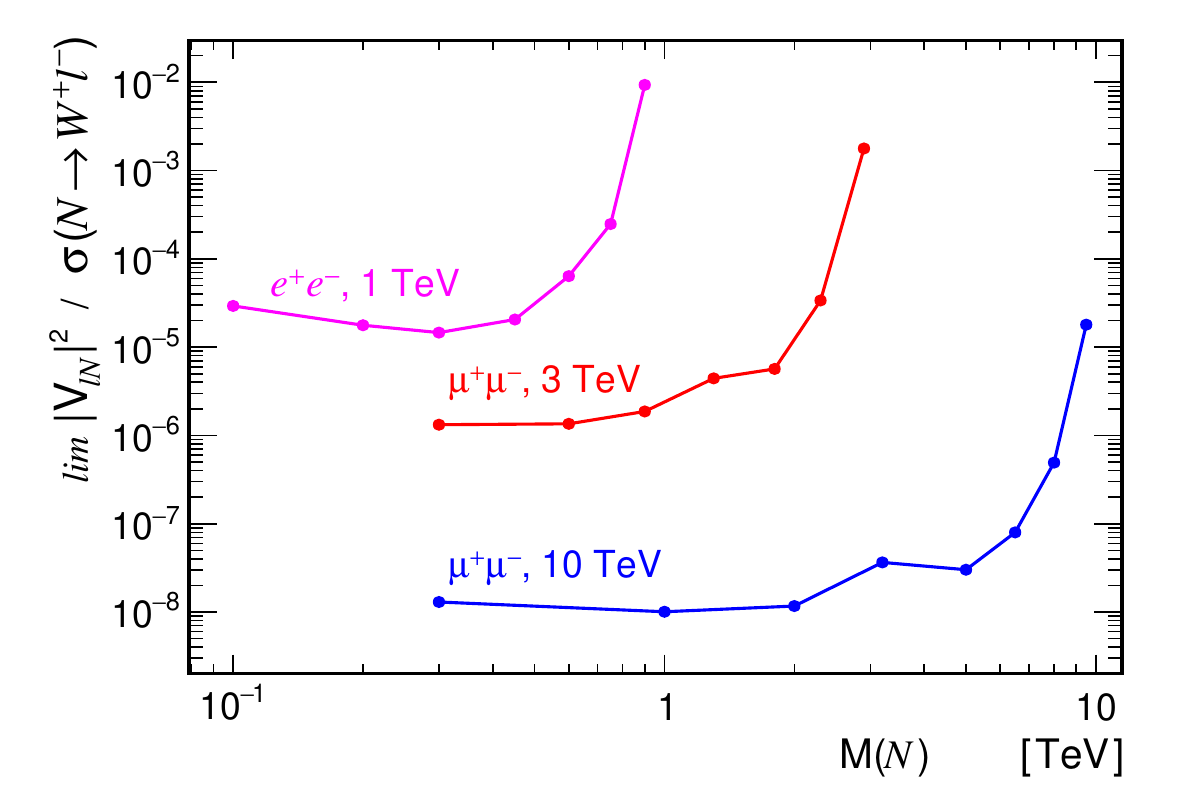}
\caption{The upper limits on the mixing parameters $|V_{\ell N}|^2$ as a function of $M(N)$ are shown
for different HNL masses and CM energies and beams.}
\end{figure}

\section{\label{sec:intro}Conclusions}

Using the model-independent approach for an analysis of SM lepton sector extension by Majorana neutrinos,
the process $\ell^+\ell^- \to N W^{\pm} \ell^{\mp}$ is studied and the upper limits on the mixing
parameters $|V_{\ell N}|^2$ are obtained as a function of $M(N)$
for different HNL masses and CM energies and beams.
The studied process can provide very competitive upper limits, especially at the multi-TeV CM energies.

Similar processes can be used to obtain upper limits also in case of the same-sign beams,
in particular in the process $\mu^+\mu^+ \to N W^+ \mu^+$. In case of the same-sign
beams the signal cross section and event kinematics are exactly the same, resulting in the
same number of the signal events. However the backgrounds are expected to be very small,
that should provide slightly better upper limits on the mixing parameters.\\

{\bf Acknowledgment \hskip 1mm} We thank A.~F.~\.Zarnecki and D.~Jeans for
valuable comments.
The work of M.D. was supported by the Russian Science Foundation Grant No.~22-12-00152.

\end{document}